# Plasmonic Meta-Surface for Efficient Ultra-Short Pulse Laser-Driven Particle Acceleration

Doron Bar-Lev[1,2] and Jacob Scheuer[1,2],*

[1]*School of Electrical Engineering Tel Aviv University, Ramat Aviv, Tel-Aviv 69978, Israel*
[2]*Tel Aviv University Center for Nano-Science and Nanotechnology, Ramat Aviv, Tel-Aviv 69978, Israel*
e-mail address: kobys@eng.tau.ac.il

A laser-driven particle accelerator based on plasmonic nano-antennas is proposed and analyzed. The concept utilizes the enhancement and localization of the electric field by nano-antennas to maximize the acceleration gradient and to overcome potential metallic losses. The structure is optimized for accelerating relativistic particles using a femto-second laser source operating at 800nm, and is shown to support the bandwidth of ultra-short laser pulses (up to 16fsec) while providing a high acceleration gradient potentially reaching 11.6GV/m.

## I. INTRODUCTION

Laser-driven particle accelerators constitute a major step towards compact and cost effective devices for high-energy particle beams generation, and thus hold promise for a variety of applications for high-energy physics, radiation generation, medicine and more. These devices utilize the high electric field of a laser beam to accelerate charged particles, where in order to attain net acceleration the oscillations of the longitudinal electric field and the position of the particles beam must be synchronized [1]. Due to its transverse nature, the electric field of a laser pulse co-propagating with the particles cannot be utilized for their acceleration [2]. Consequently, several techniques have been suggested to rotate the electric field and gain a synchronized longitudinal component that can be used [2-12]. Nonetheless, natural group velocity dispersion, relatively large mode volume and many-cycle field-buildup times render these schemes less efficient for ultra-short pulse laser utilization [13]. As such lasers are characterized by a very large electric field, exploiting them can be highly efficient. A different approach, which was recently demonstrated experimentally, utilizes a single-pass dielectric laser accelerator based on periodic field reversal [13,14]. The non-resonant nature of this structure supports ultra-short pulse operation. Nevertheless, because of its relatively low Q, the attained average longitudinal field (acceleration gradient) in a quartz based structure is $G=0.49E_0$, where $E_0$ is the electric field amplitude of the input plane wave. A principle question to be asked is whether it is possible to design a laser accelerator which provides a larger acceleration gradient while supporting ultra-short laser operation.

One of the common conceptions in laser accelerators design is that losses at optical wavelengths render metallic based devices inefficient [2,12]. However, a metallic structure exhibiting a plasmonic resonance lead to field enhancement in the near-field zone of the structure [15] which can be utilized for compensating these losses and attaining a substantial acceleration gradient. Nano-antennas are nanometer scale metallic objects which interact with electro-magnetic (EM) waves and can be designed to exhibit localized plasmon resonances at optical and infra-red frequencies. These nano-structures were shown to enhance the efficiency of numerous optical phenomena such as photodetection, photovoltaics, photogeneration, nonlinear optical phenomena, imaging and more [16]. A nano-antenna can concentrate the EM energy into an ultra-small volume [17], resulting in a large enhancement of the local field [18-21] which is attributed to the plasmonic resonances [22]. Consequently, power enhancement factors exceeding 40dB have been demonstrated [23]. Moreover, close to resonance plasmonic modes exhibit strong coupling to radiating modes which renders the impact of metallic losses less significant [24]. Similar to their conventional RF counterparts, the resonances of nano-antennas are related to their dimensions [25]. In addition, when considering nano-antenna arrays, the coupling between adjacent elements [26], the structure surroundings [27,28] and the array properties [29], can substantially affect the field enhancement. Thus, the amplitude, phase and polarization of the field in the vicinity of plasmonic meta-surfaces comprising nano-antennas can be controlled as desired [30,31]. Here, we show that plasmonic meta-surfaces can be applied for efficient laser acceleration suitable for ultra-short laser operation. In the proposed meta-surface laser accelerator (MLA) the electric field is enhanced by an array of metallic nano-antennas and is directed towards the vacuum channel (particles channel) thus enhancing the acceleration gradient. In addition, the structure is designed such that the field is mostly localized in the vacuum and it yields a specific spatial phase profile of the field which enhances the gradient further. Consequently, a large acceleration gradient bound is attained despite the use of metals. The proposed structure is designed for accelerating relativistic electrons (i.e. $\beta\approx 1$).

Nevertheless, nano-antenna arrays versatility enables the design of MLAs which are optimized for slower particle beams as well.

The rest of the paper is organized as follows: Section II describes the proposed MLA structure and presents the accelerating field distribution and the resulting acceleration gradient. The dependency of this acceleration gradient on specific design parameters as the vacuum channel width and the Au layer thickness is discussed in section III. In section IV the expected damage threshold of the structure is analyzed and the corresponding maximal unloaded acceleration gradient bound is deduced. Following the loaded gradient and bunch charge limit are also calculated. In section V we address the ultra-short laser operation and infer the shortest supported pulse. Finally in section VI we summarize the main findings and conclude.

## II. MLA STRUCTURE AND ACCELERATING FIELD DISTRIBUTION

Attaining effective acceleration requires the synchronization of the particles beam with the electric field component in the direction of the particles motion (denoted here as the *y* direction). Consequently, for $\beta \approx 1$ electron beams it is essential that the spatial distribution of the field includes a spatial harmonic with a wave number equals to $k_0 = 2\pi/\lambda_0$, where $\lambda_0$ is the center wavelength [2]. A possible method to satisfy this requirement is to introduce a periodic structure with a longitudinal period of $d_y = \lambda_0$. Let us assume a slab like structure which encloses the particles channel and is symmetric along the *z*-axis. The structure is illuminated symmetrically by two incoming fields which propagate in opposite directions along the *z*-axis and are polarized along the particle channel (see Fig. 1). The electric field in the channel can be expressed in the Floquet form, $E_y = E_0(e^{i(kz-\omega t)} + e^{-i(kz+\omega t)}) \cdot U(y)$, where $E_0$ is the amplitude of each one of the impinging fields and $U(y)$ is a periodic function $U(y+\lambda_0) = U(y)$. Consequently, it is possible to expand the field by its Fourier series:

$$E(y,z,t) = E_0\left(e^{i(kz-wt)} + e^{-i(kz+wt)}\right)\sum_n a_n e^{i\frac{2\pi}{\lambda_0}ny} \quad (1)$$

By assuming that the electrons are traveling along $z=0$ and applying the propagation relation $t \approx y/c$, the field along the trajectory can be expressed as a function of *y* only:

$$E(y) = 2E_0 e^{-i\frac{2\pi}{\lambda_0}y} \sum_n a_n e^{i\frac{2\pi}{\lambda_0}ny} \quad (2)$$

The acceleration gradient *G* is defined as the average field along a single period [13], yielding:

$$G = \frac{2E_0}{\lambda_0}\sum_n \int_0^{\lambda_0} a_n e^{i\frac{2\pi}{\lambda_0}y(n-1)} dy = 2E_0 a_1 \quad (3)$$

As can be seen, the acceleration gradient stems from the fundamental spatial harmonic of the field. Nevertheless, due to the field enhancement induced by the nano-antennas, the component $a_1$ can be large, as the field is focused from a unit cell area to a localized point. To enhance the fundamental harmonic while retaining a relatively high field enhancement, we propose utilizing a plasmonic meta-surface which consists of a periodic arrangement of unit-cells comprising two Au nano-antennas: a slot and a patch (see Fig. 1(a)). As these nano-antennas are complementary in the sense that patches excite electric currents while slots excite "magnetic currents", one can expect that combining them into a unit cell may yield the desired electric field phase profile for effective acceleration. Clearly, the electron beam might be affected by the existence of transverse electric fields. Therefore, to maximize the acceleration and maintain electron beam quality [2] we utilize a symmetric MLA structure comprising two parallel meta-surfaces of thickness *t* which are patterned on $SiO_2$ substrates. These meta-surfaces are placed at the opposite facets of a $d_z$ wide vacuum channel as depicted in Fig. 1(b). The acceleration gradient upper bound stems from the expected damage threshold of the MLA structure. At the nano-antennas resonance the field is expected to be localized in the vicinities of the gaps of the structure. Therefore to minimize the field in the $SiO_2$ substrates and attain a large structural damage threshold, we remove the $SiO_2$ from the vicinity of the metallic-gap boundaries, creating 80nm deep mushroom-like vacancies beneath the slots and the gaps adjacent to the patches (see highlighted oval areas in Fig. 1(b) and Fig. 1(c)). Such apertures can be realized by introducing an anisotropic etching stage to the MLA fabrication process. The structure is periodic in both the *x* and *y* directions. The periodicity in *x* affects the coupling between adjacent unit cells and is optimized for obtaining (approximately) a $\pi$ phase shift in the accelerating field profile after the accelerated electron beam propagates half a unit cell, thus maximizing the acceleration gradient. The periodicity in *y* is set to be $\lambda_0$, therefore maintaining the required phase synchronicity over many MLA periods.

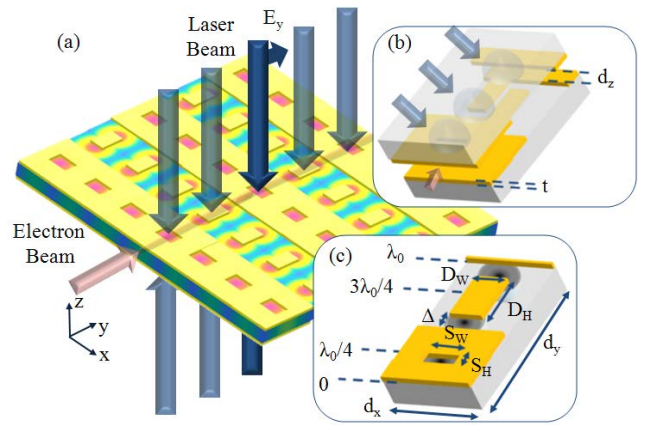

FIG. 1. MLA Schematic description; (a) Top view ; (b) Unit cell structure; (c) Unit cell dimensions.

Using a finite difference time domain (FDTD) simulation tool [32], we optimized the MLA dimensions such that it resonates at $\lambda_0=800$nm. This choice stems from the availability of high-quality, high-power laser sources at this wavelength (e.g. Ti:Sapphire lasers). Nevertheless, different resonances can be designed as well. The impinging laser fields are linearly polarized along the *y* axis, corresponding to the maximal response of the nano-antennas. In addition, symmetric illumination in the ±*z* directions is employed to eliminate the impact of transverse electric fields. Consequently, we normalize the results by $2E_0$, where $E_0$ is the amplitude of each one of the impinging fields. The analysis takes into account the measured dispersion of the employed materials as well as the optical losses in the metallic layers [33]. Specifically at 800nm the Au dielectric constant is complex - $\varepsilon=-25.43+1.83i$, which is the value employed for the analysis. Table I, details the MLA dimensions for attaining a resonant response at 800nm.

**TABLE I.** MLA dimensions [nm] for 800nm resonance

| $d_x$ | $d_y$ | $d_z$ | t | $S_w$ | $S_H$ | $D_W$ | $D_H$ | $\Delta$ |
|---|---|---|---|---|---|---|---|---|
| 500 | 800 | 100 | 40 | 200 | 100 | 150 | 147 | 75 |

As can be seen, the array spacing along the *x*-direction, $d_x$, is of sub-wavelength scale, hence no diffraction lobes are expected in this direction. The *y* spacing on the other hand corresponds to the limit of the 1st diffraction order condition in the vacuum. However, because of the $SiO_2$ substrates, the MLA structure does not suffer from losses that originate from Wood's anomalies. The remainder of the dimensions is in the order of 100nm, thus feasible by means of contemporary fabrication techniques (e.g. E-beam lithography). Fig. 2 depicts the distribution of the longitudinal electric field ($E_y$) along the symmetry axis of the MLA (*x*=*z*=0). This axis corresponds to the electron beam trajectory (highlighted line in Fig. 1(a)). Fig. 2(a) depicts the dependence of the $E_y$ amplitude on the laser wavelength and denotes the optimal acceleration field at 800nm (normalized to $2E_0$) experienced by a relativistic particle with the maximally accelerating phase. Fig. 2(b) and Fig. 2(c) respectively depict $E_y$ amplitude and phase at λ=800nm. Clearly, the maximal field is obtained around λ=800nm corresponding to the MLA resonance. The field is localized at three different regions in which the normalized amplitude ($|E_y|/2E_0$) is enhanced by a factor ranging between 6 to 3 (see Fig. 2(b)). The first region corresponds to the area between the slots openings (*y*=-$\lambda_0/4$=-0.2μm), while the other two are located symmetrically around *y*=$\lambda_0/4$=0.2μm, i.e. at the apertures adjacent to the patches. The phase difference between the first and the two other peaks is $0.83\pi$ (see Fig. 2(c)), resulting primarily in an acceleration buildup along the electron beam trajectory. Note that there is also a small peak at *y*=*0.2*μm with approximately the same phase as that of the slot antenna peak. This peak lowers the acceleration gradient however it is substantially smaller than the other peaks, therefore has low impact.

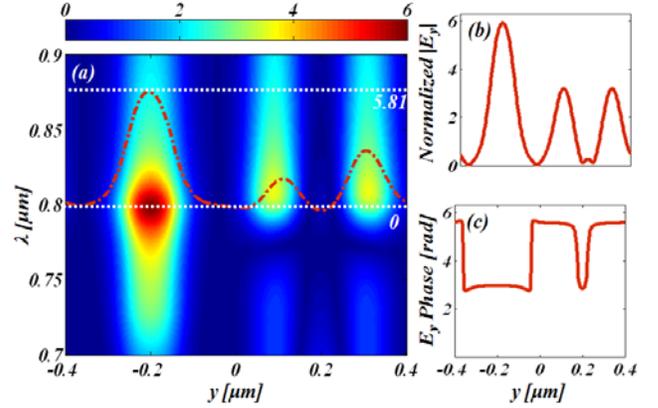

FIG. 2. Longitudinal electric field distribution. (a) Normalized amplitude wavelength dependency, dashed line - optimal acceleration field at 800nm; (b) Normalized amplitude at 800nm; (c) Phase at 800nm.

The acceleration gradient is evaluated using Eq. (3), yielding a normalized gradient of $G/2E_0=1.32$ for the structure parameters described in Table I. Note that at a given laser beam power, splitting the beam for symmetrical illumination decreases the amplitude of the electric-field from each side by a factor of $\sqrt{2}$, however it increases the acceleration gradient by the same factor. Hence symmetrical illumination yields a more efficient acceleration. Therefore, when comparing the acceleration gradient to that provided by the dielectric field reversal structure presented in Ref. [13], the MLA attains approximately a 3.81 times larger gradient.

### III. VACUUM CHANNEL WIDTH AND AU LAYER THICKNESS OPTIMIZATION

The MLA parameters affect the resonance wavelength, the field enhancement, and the electric field phase distribution along the vacuum channel. Consequently, each of these parameters affects the acceleration gradient. Fig. 3 shows the impact of the electron channel width, $d_z$ (Fig. 3(a)) and the Au layer thickness, *t* (Fig. 3(b)) on the normalized acceleration gradient. The gradient increases monotonically for smaller $d_z$, reaching a gradient of $3.35E_0$ for $d_z$=40nm. This is a direct result of the field enhancement and localization properties induced by the nano-antennas. The field localization points where the field enhancement is maximal are located at the gaps of the slots and at the apertures adjacent to the dipoles, approximately at the mid thickness of the Au layer. Therefore, smaller $d_z$ results in stronger coupling of these intense fields to the vacuum channel, consequently increasing the acceleration gradient. However, for practical reasons, it is desired to employ the widest possible vacuum channel. Hence, a 100nm width is a good compromise as it yields a relatively large gradient while retaining a sufficiently wide vacuum channel. Note that at a width of $\lambda_0/4$=200nm (the optimum width of the dielectric field reversal accelerator) the acceleration gradient equals $1.03E_0$ (assuming equivalent laser beam power and double illumination). This value still exceeds the $0.49E_0$ gradient reported in ref. [13].

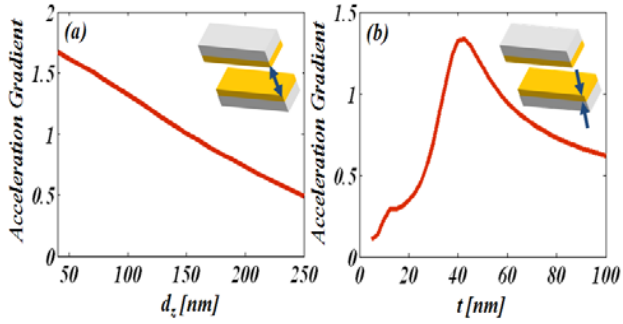

FIG. 3. Normalized acceleration gradient as a function of: (a) vacuum channel width; (b) Au layers thickness. All other parameters correspond to the optimized MLA for 800nm resonance.

The Au thickness affects the resonance wavelength and can therefore modify the acceleration gradient as well. This effect is clearly seen in Fig. 3(b) exhibiting a distinctive peak at $t=40$nm, for which the resonance wavelength is 800nm. Note that for thick layers the field enhancement is mostly determined by the Au-SiO$_2$ surface and is thus less sensitive to the thickness variations of the Au layer

## IV. ACCELERATION GRADIENT BOUND, LOADED GRADIENT AND BUNCH CHARGE LIMIT

To deduce the MLA acceleration gradient bound an investigation of the expected structural damage is required. Structural damage under laser illumination is caused mainly due to melted heat zones and laser-ablation, where for ultra-short pulse laser the latter is the dominant effect [34-36]. While some metals, as Cu, experience lower damage thresholds than SiO$_2$ and other dielectrics [37,38], thin Au films on dielectric substrates can attain similar and even larger damage thresholds from that of SiO$_2$ when exposed to fsec light pulses. This depends on the exact pulse characteristics, the dielectric substrate and the fabrication surface quality [36,39,40]. Therefore, we assume a representative damage threshold similar to that of SiO$_2$ (~2J/cm$^2$ for a 100fsec pulse at 800nm [37-39] corresponding to a peak electric field of $E_0$~10GV/m). To deduce the acceleration gradient bound, we investigated the MLA field magnitude distribution at resonance. Fig. 4 describes the yz cross section of this distribution (normalized to $E_0$), where the white lines represent the MLA contour. As can be seen, the field is localized in the gaps, between the metallic sections. Moreover, due to the mushroom like apertures the field in the SiO$_2$ is relatively small. The maximal value is experienced at the surfaces between the different media where it reaches a maximal value of $5.69E_0$ in the Au (at the patch edges) and $3.88E_0$ in the SiO$_2$ (at the mushroom edges). Note that the field drops rapidly inside the media.

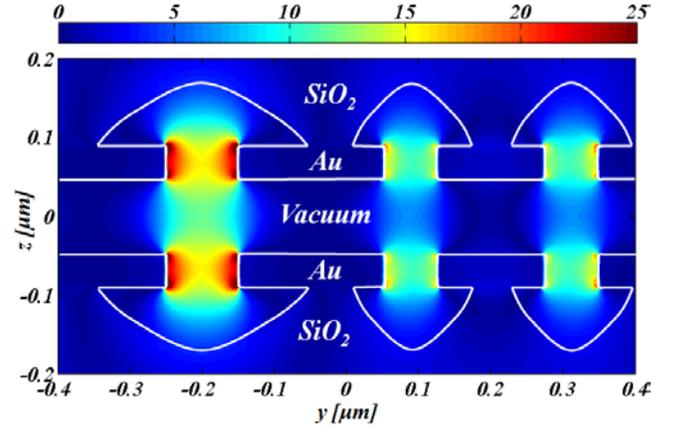

FIG. 4. A yz cross-section of the field magnitude distribution of the MLA structure for double illumination. The white lines represent the structural contour.

Consequently, we estimate the MLA maximal unloaded acceleration gradient to be 4.64GV/m for 100fsec pulses. When applying the shortest supported pulse (16fsec-see following analysis) the MLA bound is increased by a factor of 2.5 up to 11.6GV/m. Note that without the mushroom vacancies, large electric fields would form in the SiO$_2$ substrates, leading to a substantially lower bound (~0.6GV/m for a 100fsec pulse).

These relatively large bounds render the MLA highly power efficient. However this may limit the amount of charge that can be accelerated due to the wake-fields of the bunch charge. These wake-fields alter the acceleration gradient as EM fields moving with the relativistic particles encounter geometric variations of the MLA structure, get scattered and act on trailing particles in the same or consecutive bunch [41]. To explore this effect and deduce a bunch charge limit for the MLA structure, let us first note that the MLA supports the acceleration of only one bunch per laser pulse, as consecutive bunches which are located on different points cannot all remain synchronized with the accelerating field. As the impinging laser is perpendicular to the particle beams, electric fields co-propagating with the particle beam are negligible. Moreover, the stored EM energy in the structure is small as the structure attains a relatively small Q which stems only from the nano-antenna resonance (and not from a cavity like resonance between the two MLA facets) and a small reactive impedance due to the relatively wide vacuum channel. Hence, the model that was applied for the dielectric field reversal structure [13] also applies here, with the exception that the double illumination and different acceleration gradient should be considered. Under double illumination the particle experiences a laser field of $2E_0$. Therefore, in this case, by applying energy balance considerations and assuming that the non-overlapping wake-fields are small, the energy gained by a single particle passing a single unit cell can be calculated by:

$$qV = -\frac{A \cdot \lambda_0}{Z_0 \cdot c}(E_w^2 + 2E_0 E_w) \quad (4)$$

where $E_w$ is the wake-field of the particle which overlaps the applied laser fields and $2E_0$ is the laser field which the particle experience. $Z_0=377\Omega$ is the vacuum impedance and $A=D\cdot\lambda_0$ is the area (in the xy plane) which is illuminated by the field, where $D$ is the width of the impinging pulse in the x direction. . Consequently, by assuming that the wake-field is substantially smaller than the laser field and is proportional to it, the loaded gradient for a bunch with $N_b$ particles can be described in the MLA case under double illumination by:

$$G_L = G - \frac{N_b q c}{4\lambda_0^2} Z_S \quad (5)$$

where $G$ is the unloaded gradient, $q$ is the particle charge and $Z_S$ is characteristic impedance of the structure. $Z_S$ is defined as the square of the voltage gained over the power lost in the structure [41], hence by assuming small nano-antennas losses [24] it can be described by:

$$Z_S = \frac{<G>^2 \lambda_0}{D} Z_0 \quad (6)$$

where $<G>$ is the normalized acceleration gradient which corresponds to 2.64 in the MLA. Taking $D$ to be equal to $10d_x=5\mu m$, which is sufficiently wide to attain the required accelerating field profile, the characteristic impedance of the MLA equals $420.4\Omega$, larger than the $37\Omega$ impedance of the dielectric field reversal structure [13] due to the larger acceleration gradient and the double illumination. The efficiency $\eta$ is defined as the ratio between the energy gained by the accelerated particle and the applied laser energy. Therefore, without loss of generality, for a unit cell length of $\lambda_0$ and corresponding time $\lambda_0/c$ during which the particle is affected by the pulse in this unit cell, the efficiency is related to the bunch charge as follows:

$$\eta = \frac{N_b q c G_L}{P_L} \quad (7)$$

where $P_L = \frac{AE_0^2}{2} \cdot 2$ is the applied laser power that is required for a double illumination of amplitude $E_0$. By introducing Eq. (5) into Eq. (7) it can be seen that there is a quadratic relation between $\eta$ and $N_b$. Hence $\eta$ has an optimum which can be derived by deducing the extremum of this relation. At this optimum efficiency $G_L=0.5G$ and the charge limit is:

$$N_{b\_opt} = \frac{2G\lambda_0^2}{qcZ_s} \quad (8)$$

At the bound acceleration gradient for 100fsec the charge limit is thus $N_{b\_opt}\sim 2.94\cdot10^5$ (assuming electron acceleration). This charge limit is significantly larger than the charge per bunch supported by waveguide based geometries [9]. However it is small compared to RF-technology based accelerators. Nevertheless, high particle beam luminosities can be yet attained by applying small beam diameters and working with high bunch and laser repetition rates. Note that the MLA charge limit is approximately an order of magnitude smaller than the dielectric field reversal bunch charge limit ($\sim 3\cdot10^6$). This is due to its larger acceleration gradient and relatively large characteristic impedance.

## V. TEMPORAL LIMITATIONS AND SHORTEST SUPPORTED PULSE

The effective bandwidth of the MLA determines its compatibility to ultra-short pulse laser operation. The pass-band of the MLA stems from the resonance linewidth of the nano-antennas and from the accumulated de-synchronization which occurs when the wavelength of the incident field differs from the longitudinal structure periodicity ($d_y=\lambda_0$). The solid line in Fig. 5(a) depicts the calculated spectral bandwidth (full width half maximum - FWHM) of the MLA structure detailed in Table I as a function of the number of periods. As expected, it decreases monotonically for larger number of periods due to the de-synchronization effect, where the width at N=1 corresponds to that of the resonance linewidth (68nm). To evaluate the shortest temporal pulse supported by the MLA structure, we compare the FWHM bandwidth of a transform limited Gaussian temporal pulse to the pass-band of the MLA. Assuming a center wavelength of $\lambda_0$ and a $\tau_p$ FWHM temporal width, the pulse bandwidth is given by [42]:

$$\Delta\lambda = \frac{\lambda_0^2}{c} \cdot \frac{2\ln(2)}{\pi} \cdot \frac{1}{\tau_p} \quad (9)$$

The electrons passing through the MLA are accelerated only when a laser pulse exists, i.e. along a path length of $\tau_p \cdot c$. This length corresponds to $N_{period}=\tau_p \cdot c/\lambda_0$ number of consecutive periods. By introducing this relation into Eq. (9), the pulse bandwidth can be described as a function of the corresponding number of structure periods:

$$\Delta\lambda = \frac{2\ln(2)}{\pi} \cdot \frac{\lambda_0}{N_{period}} \quad (10)$$

The dashed line in Fig. 5(a) describes the pulse bandwidth as a function of the number of periods that is occupies.

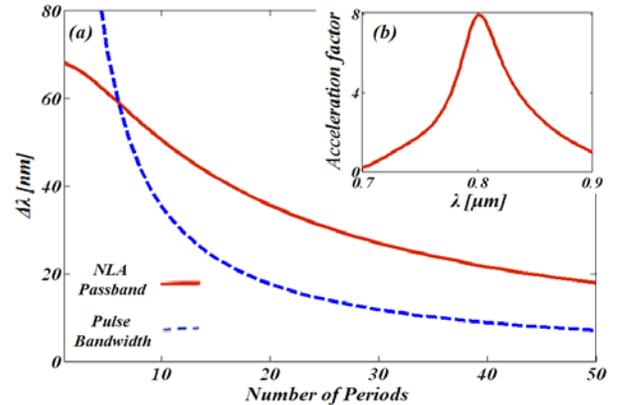

FIG. 5. (a) MLA pass-band (solid red) and pulse bandwidth (dashed blue) as a function of the MLA number of periods; (b) Normalized acceleration factor for ten periods long MLA.

As can be seen this bandwidth also decreases monotonically with increasing number of periods. Nonetheless, when the pulse occupies more than six periods (a pulse larger than 16fsec) the pass-band of the MLA becomes larger than the pulse bandwidth. Therefore, we conclude that the MLA supports excitation pulses which are longer than 16fsec (note that CW operation is also supported). The total normalized acceleration factor, i.e. $G \cdot N_{period}/2E_0$, corresponding to the crossing point is depicted in Fig. 5(b). As can be seen the response is almost symmetrical, centered at the resonance wavelength, $\lambda_0=800nm$.

## VI. CONCLUSIONS

The MLA structure proposed here constitutes a new paradigm for laser-driven particle acceleration. It utilizes the strong field enhancement and localization provided by a meta-surface comprising periodic slot-patch metallic nano-antenna array, to attain net acceleration. This approach yields a relatively large acceleration gradient, while supporting ultra-short laser operation. The proposed MLA design was optimized for relativistic particle beams and laser operation at λ=800nm. Under symmetrical illumination it provides an acceleration gradient of $G=2.64E_0$, which is 3.81 times larger than the expected maximal acceleration gradient of the dielectric field reversal structure. The symmetric illumination contributes a factor of $\sqrt{2}$ to this ratio. Elaborating on this observation, an MLA enabling illumination from all four facets of the vacuum channel can double the efficiency and an MLA with cylindrical symmetry which utilizes radial illumination seems the most efficient. Nevertheless, these structures are more complicated to realize. The presented structure is applicable for ultra-short laser operation up to a lower bound of 16fsec. The maximal acceleration gradient at this bound is estimated as 11.6GV/m. This estimation is based on empirical damage thresholds and on calculations of the maximal field magnitude in the media. The latter values were reached by the introduction of the mushroom like vacancies which reduce the maximal field in the substrates. We believe that the proposed MLA opens new research directions for additional plasmonic based accelerators. Particularly, the large number of structural degrees of freedom facilitates the design of MLAs operating at different laser center wavelengths and for slower particle beams. The main challenge in designing these structures is attaining a field profile that resembles the required spatial harmonic as slower particles acceleration necessitates smaller unit cells. The MLA design versatility along with the efficient ultra-short pulse operation and the relatively high acceleration gradient, render the MLA a promising concept for future laser-driven particle accelerators operating at the ultra-short pulse regime.

## ACKNOWLEDGMENTS

The authors want to thank A. Gover for thorough and stimulating discussions regarding the MLA concept and its analysis. D. B. acknowledges the generous support of the Tel-Aviv University center for Nano-Science and Nanotechnology.